\documentclass[11pt]{article}
\newcommand{\Comment}[1]{{}}
\usepackage[textwidth = 430 pt, textheight = 630 pt]{geometry}
\usepackage{amssymb,euscript,amsmath,amsfonts}

\usepackage{color}
\definecolor{MyDarkBlue}{rgb}{0.15,0.15,0.45}
\usepackage[linktocpage=true]{hyperref}
\hypersetup{
colorlinks=true,
citecolor=MyDarkBlue,
linkcolor=MyDarkBlue,
urlcolor=MyDarkBlue,
pdfauthor={Neil Lambert and Constantinos Papageorgakis},
pdftitle={},
pdfsubject={hep-th}
}

\usepackage[numbers,sort&compress]{natbib}
\usepackage{hypernat}

\newcommand\sfrac[2]{{\textstyle\frac{#1}{#2}}}
\newcommand\ignore[1]{}
\def\one{{\,\hbox{1\kern-.8mm l}}}

\def\Tr{{\rm Tr\, }}

\newcommand{\SO}{\mathrm{SO}} 
\newcommand{\SU}{\mathrm{SU}} 
 \newcommand{\pd}{\partial}

\newcommand{\Cset}{{\,\,{{{^{_{\pmb{\mid}}}}\kern-.45em{\mathrm C}}}}}

\newcommand{\nn}{\nonumber}
\newcommand{\ie}{{\it i.e.}}
\newcommand{\eg}{{\it e.g.}}
\newcommand{\be}{\begin{equation}}
\newcommand{\ee}{\end{equation}}
\newcommand{\bea}{\begin{eqnarray}}
\newcommand{\eea}{\end{eqnarray}}

\begin{document}

\renewcommand{\thefootnote}{\fnsymbol{footnote}}

\rightline{}
   \vspace{1.8truecm}

\vspace{15pt}

\centerline{\LARGE \bf {\sc Nonabelian $(2,0)$ Tensor Multiplets }} \vspace{.5cm}
\centerline{\LARGE \bf {\sc and 3-algebras}}
\vspace{2truecm}
\thispagestyle{empty}
\centerline{
    {\large {\bf {\sc Neil Lambert${}^{\,a,}$}}}\footnote{On leave of absence from King's College London.}$^,$\footnote{E-mail address: \href{mailto:neil.lambert@cern.ch}{\tt neil.lambert@cern.ch}} { and}
    {\large {\bf{\sc Constantinos Papageorgakis${}^{\,b,}$}}}\footnote{E-mail address:
                                 \href{mailto:costis.papageorgakis@kcl.ac.uk}{\tt costis.papageorgakis@kcl.ac.uk}}
                                                           }

\vspace{1cm}
\centerline{${}^a${\it Theory Division, CERN}}
\centerline{{\it 1211 Geneva 23, Switzerland}}
\vspace{.8cm}
\centerline{${}^b${\it Department of Mathematics, King's College London}}
\centerline{{\it The Strand, London WC2R 2LS, UK}}

\vspace{2.0truecm}

\thispagestyle{empty}

\centerline{\sc Abstract}

\vspace{0.4truecm}
\begin{center}
\begin{minipage}[c]{380pt}{
\noindent 
Using 3-algebras we obtain a nonabelian system of equations that furnish a representation of the  $(2,0)$-supersymmetric  tensor multiplet. The on-shell conditions are quite restrictive so that the system can be reduced to five-dimensional gauge theory along with  six-dimensional abelian $(2,0)$ tensor multiplets. We briefly discuss possible applications to D4-branes using a spacelike reduction and M5-branes using a null reduction.
}
\end{minipage}
\end{center}

\vspace{.4truecm}

\noindent

\vspace{.5cm}

\setcounter{page}{0}

\newpage

\renewcommand{\thefootnote}{\arabic{footnote}}
\setcounter{footnote}{0}

\linespread{1.1}
\parskip 4pt

{}~
{}~

\makeatletter
\@addtoreset{equation}{section}
\makeatother
\renewcommand{\theequation}{\thesection.\arabic{equation}}

\section{Introduction}

There has been significant recent progress in the formulation of  Lagrangian descriptions for multiple M2-branes in M-theory \cite{Bagger:2006sk,Bagger:2007jr,Bagger:2007vi,Gustavsson:2007vu,Aharony:2008ug,Bagger:2008se}. These descriptions initially relied on the introduction of a novel algebraic structure, going under the name of a 3-algebra. Even though one can recast the former in terms of a completely conventional gauge theory language, the presence of  3-algebras is intriguing and one might wonder about their possibly deeper connections to M-theory in general. 

In this note we begin the investigation of a potential relation between 3-algebras and multiple  M-theory fivebranes. Compared to M2-brane systems the formulation of an M5-brane theory is difficult at best: Even for the case of a single fivebrane it does not seem possible to write down a six-dimensional action with conformal symmetry due to the selfduality of the three-form field-strength. In addition the theory of multiple M5-branes is given by a conformal field theory in six-dimensions with  mutually local electric and magnetic states  and no coupling constant. All of these features are difficult to reconcile with  a Lagrangian description.\footnote{For a review on M2 and M5-brane basics see \cite{Berman:2007bv}.}

Here we will simply study the equations of motion of a nonabelian $(2,0)$ tensor multiplet. Starting with the set of supersymmetry transformations for the abelian M5-brane we propose an ansatz for a nonabelian generalisation.  Apart from the expected nonabelian versions of the scalars, fermions and the antisymmetric three-form field strength, we introduce a gauge field as well as a non-propagating  vector field which  transforms nontrivially under the nonabelian gauge symmetry and has a negative scaling dimension.  Our ansatz involves `structure constants' with four indices that can be associated to a 3-algebra.

It is interesting to note that there have been proposals for M5-brane Lagrangians that require the introduction of a new scalar field, which can be eliminated at the price of sacrificing manifest six-dimensional Lorentz invariance \cite{Pasti:1997gx,Bandos:1997ui}. Although here we will only study the equations of motion, the additional non-propagating vector field that we introduce is a nonabelian analogue of the auxiliary field in  \cite{Pasti:1997gx,Bandos:1997ui}. For recent work that also touch upon some of these issues  see \cite{Chu:2009iv,Huddleston:2010cx,Chen:2010br,Rey:2010uz,Berman:2009xd}.

We proceed by studying the closure of the supersymmetry algebra. We find that it closes on-shell up to translations and gauge transformations if the structure constants are those of real 3-algebras,\footnote{These are the $\mathcal N=8$ 3-algebras in three dimensions.} \ie~they are totally antisymmetric and obey the related `fundamental identity'. The on-shell conditions yield a set of equations of motion for the various fields, as well as a number of constraints. The latter prove to be quite restrictive and upon expanding the theory around a particular vacuum it reduces to five-dimensional super-Yang-Mills along with six-dimensional, abelian $(2,0)$ tensor multiplets. In particular we will essentially arrive at a reformulation of the D4-brane theory with conformal symmetry and $(2,0)$ supersymmetry. This is similar to the Lorentzian M2-brane models \cite{Gomis:2008uv,Benvenuti:2008bt,Ho:2008ei} which provided a different angle on D2-branes \cite{Ezhuthachan:2008ch}. In this way we hope that new light can be shed on M5-branes by reformulating D4-branes in terms of a $(2,0)$ system. In addition this paper can be viewed as a no-go theorem for obtaining a genuine six-dimensional interacting $(2,0)$ supersymmetric set of equations of motion. On the other hand a different null reduction leads to a novel system with 4 space and 1 null directions, sporting lightlike dyonic-instanton string BPS solutions \cite{Lambert:1999ua}. We conclude with some further remarks on the possible connection of this theory to the dynamics of multiple M5-branes.

\section{A nonabelian $(2,0)$ Tensor Multiplet}

We start by giving the covariant supersymmetry transformations of a free six-dimensional $(2,0)$ tensor multiplet \cite{Howe:1997fb}:
\bea\label{Habelian}
\delta X^I &=& i \bar \epsilon \Gamma^I \Psi\cr
\delta  \Psi &=& \Gamma^\mu \Gamma^I \pd_\mu X^I \epsilon + \frac{1}{  3!}\frac{1}{2} \Gamma^{\mu\nu\lambda}H_{\mu\nu\lambda}\epsilon\; \cr
\delta B_{\mu\nu} &=& i \bar \epsilon \Gamma_{\mu\nu} \Psi \; ,
\eea
where $\mu = 0,...,5$, $I = 6,...,10$ and $H_{\mu\nu\lambda} = 3\partial_{[\mu}B_{\nu\lambda]}$ is selfdual. The supersymmetry generator $\epsilon$ is chiral: $\Gamma_{012345}\epsilon=\epsilon$ and the Fermions $\Psi$ are antichiral: $\Gamma_{012345}\Psi=-\Psi$.
This algebra closes on-shell with the equations of motion
\bea
\Gamma^\mu \pd_\mu \Psi =\pd_\mu\pd^\mu X^I =0\; .
\eea 
We note that, from the point of view of supersymmetry, it is sufficient to write the algebra purely in terms of $H_{\mu\nu\lambda}$ and not mention $B_{\mu\nu}$:
\bea\label{Habelian}
\delta X^I &=& i \bar \epsilon \Gamma^I \Psi\cr
\delta  \Psi &=& \Gamma^\mu \Gamma^I \pd_\mu X^I \epsilon + \frac{1}{  3!}\frac{1}{2} \Gamma^{\mu\nu\lambda}H_{\mu\nu\lambda}\epsilon\; \cr
\delta H_{\mu\nu\lambda} &=& 3i \bar \epsilon \Gamma_{[\mu\nu}\partial_{\lambda]} \Psi \; ,
\eea
in which case one must include the equation of motion $\pd_{[\mu}H_{\nu\lambda\rho]}=
0$.

We wish to try and generalise this algebra to allow for nonabelian fields and interactions. To this end we assume all fields take values in some vector space with a basis $T_A$, {\it viz}. $X^I = X^I_AT^A$, {\it etc.}, and promote the derivatives to suitable covariant derivatives
\bea\label{deriv}
D_\mu X^I_A = \pd_\mu X^I_A - \tilde A_\mu^B{}_AX^I_B\;\ , 
\eea
where $\tilde A_\mu^B{}_A$ is a gauge field. 

Upon reduction on a circle one expects that the six-dimensional  $(2,0)$ transformation rules reduce to those of the five-dimensional super-Yang-Mills, which are given by
\bea\label{SYM}
\delta X^I &=& i \bar \epsilon \Gamma^I \Psi\nn\\
\delta  \Psi &=& \Gamma^\alpha \Gamma^I D_\alpha X^I \epsilon + \frac{1}{  2} \Gamma^{\alpha\beta}\Gamma^{5}F_{\alpha\beta}\epsilon
-\frac{i}{ 2}[X^I,X^J]\Gamma^{IJ}\Gamma^{5}\epsilon\; \nn\\
\delta A_{\alpha} &=& i \bar \epsilon \Gamma_{\alpha}\Gamma_{5} \Psi \;,
\eea
for $\alpha = 0,...,4$.

In order to obtain a term analogous to the $[X^I,X^J]$ for $\delta\Psi$ in (\ref{Habelian}) we need to introduce a $\Gamma_\mu$ matrix to account for the fact that $\epsilon$ and $\Psi$ have opposite chirality. Thus a natural guess is to  propose the existence of a new field $C^\mu$ so that we can consider the  ansatz:
\bea\label{ansatz1}
\delta X^I_A &=& i \bar \epsilon \Gamma^I \Psi_A\cr
\delta \Psi_A &=& \Gamma^\mu \Gamma^I D_\mu X_A^I \epsilon + \frac{1}{ 3!}\frac{1}{2} \Gamma_{\mu\nu\lambda}H_A^{\mu\nu\lambda}\epsilon- \frac{1}{2}\Gamma_\lambda \Gamma^{IJ} C^\lambda_B X^I_C X^J_D {f^{CDB}}_A\epsilon\cr
\delta H_{\mu\nu\lambda\; A} &=& 3 i \bar \epsilon \Gamma_{[\mu\nu}D_{\lambda]} \Psi_A +  i\bar \epsilon \Gamma^I \Gamma_{\mu\nu\lambda\kappa}C^\kappa_{B} X^I_C \Psi_D{g^{CDB}}_A \cr
\delta  \tilde A_{\mu\;A}^{\;B} &=& i \bar \epsilon \Gamma_{\mu\lambda} C^\lambda_C \Psi_D {h^{CDB}}_A\cr
\delta C^\mu_A  &=& 0\;,
\eea
Here $f^{CDB}{}_A$,  $g^{CDB}{}_A$ and  $h^{CDB}{}_A$ are `structure' constants that we will determine in due course. Note that we can assume $f^{CDB}{}_A$ is antisymmetric in $C,D$.

As with the abelian case we also impose selfduality on $H_{\mu\nu\lambda A}$:
\bea
H_{\mu\nu\lambda A} = \frac{1}{3!}\epsilon_{\mu\nu\lambda\tau\sigma\rho}H^{\tau\sigma\rho}{}_A\;.
\eea 
Demanding that this is preserved under supersymmetry gives rise to the Fermion equation:
\be\label{fermionanticipated}
\Gamma^\lambda D_\lambda \Psi_A + C_{\lambda\;B }X^I_C\Gamma^\lambda\Gamma^I \Psi_D{g^{CDB}}_A  =0\;.
\ee

Note that consistency of the above set of equations with respect to their scaling dimensions gives
\bea
[H] = [X]+1 \;,\qquad &&\qquad [\tilde A] = 1\;,\qquad\qquad [C] = 1-[X]\cr
[\epsilon] = -\sfrac{1}{2}\;,\qquad &&\qquad [\Psi] = [X] +\sfrac{1}{2}\;,\qquad\qquad [X] \;.
\eea
so one could still make this work with some other assignment  that are all related to the choice of $[X]$. However the canonical choice is
$[X]=2,[H]=3,[\Psi]=\frac{5}{2},[C]=-1$. In particular we see that the new field $C^\mu$ has scaling dimension $-1$. Therefore, if we compactify the theory on a circle of radius $R$ we expect the expectation value of $C^\mu$ to be proportional to $R$. 

\subsection{Closure on $X^I_A$}

We now proceed to test the ansatz (\ref{ansatz1}) by investigating the closure of the supersymmetry algebra  on the scalars. A straightforward calculation gives
\bea\label{Xs}
[\delta_1, \delta_2]X^I_A &=&  v^\mu D_\mu X^I_A + \tilde \Lambda^B{}_AX^I_B\;,
\eea
where
\bea
v^\mu = -2i (\bar\epsilon_2\Gamma^\mu\epsilon_1)\;,\qquad \tilde \Lambda^B{}_A = 2i(\bar \epsilon_2\Gamma_\lambda\Gamma^J\epsilon_1) C^\lambda_DX^J_C f^{BCD}{}_A\;.
\eea

\subsection{Closure on $C^\mu_A$}

Here the situation is rather simple since one clearly has $[\delta_1,\delta_2]C^\mu_A=0$. On the other hand  what we expect is
\be
[\delta_1, \delta_2]C^\mu_A =v^\nu D_\nu C^\mu_A +\tilde \Lambda^B_{\;A} C^\mu_B\;.
\ee
These agree on-shell if
\bea\label{Cconstraints}
D_\nu C^\mu_A = 0\;,\qquad  C^\lambda_B C^\rho_C {f^{CDB}}_A =0 \label{Cf}\;.
\eea
The second constraint comes from setting $\tilde \Lambda^B{}_AC^\mu_B=0$ but also comes from supersymmetrising the first constraint.

\subsection{Closure on $\tilde A^A_{\mu \;B}$}

We continue with the closure on the gauge fields. In order to minimise the size of our expressions we will freely use the constraints found above. We obtain
\bea\label{As}
[\delta_1,\delta_2]\tilde A^B_{\mu \;A} &=& - 2i(\epsilon_2\Gamma_\lambda\Gamma^I\epsilon_1) C^\lambda_C D_\mu X^I_Dh^{CDB}{}_A-v^\nu C^\lambda_C H_{\mu\nu\lambda D}h^{CDB}{}_A \cr
&&-i(\bar\epsilon_2\Gamma_{\mu\nu\lambda}\Gamma^{IJ}\epsilon_1) C^\nu_CC^\lambda_GX^I_EX^J_Ff^{EFG}{}_Dh^{CDB}{}_A\cr
&&+
2i(\epsilon_2\Gamma_\mu\Gamma^I\epsilon_1)C^\lambda_C D_\lambda X^I_D h^{CDB}{}_A\cr
&=& v^\nu \tilde F^{\;B}_{\mu\nu\; A} + D_\mu  \tilde \Lambda^B_{\;\;A}\;.
\eea
where in the last line we have given what the first three lines should amount to, and the  field-strength is defined from (\ref{deriv}) through $[D_\mu , D_\nu] \equiv  \tilde F_{\mu\nu}$:
\be\label{Fis}
 \tilde F^{\;\;A}_{\mu\nu\;B} = \pd_\nu \tilde A^{\;A}_{\mu \;B} - \pd_\mu \tilde A^{\;A}_{\nu \;B} - \tilde A_{\mu\;C}^{\;A}\tilde A_{\nu\;B}^{\;C} + \tilde A_{\nu\;C}^{\;A}\tilde A_{\mu\;B}^{\;C} \;.
\ee

For the first term to give the correct gauge transformation we deduce that
\bea
{h^{CDB}}_A &=&f^{DBC}{}_A\label{fhone}\;.
\eea
Given this we see that the second term is a translation provided that
\be\label{gaugeeom}
\tilde F^{\;B}_{\mu\nu\; A} = C^\lambda_C H_{\mu\nu\lambda\;D}{f^{BDC}}_A \;.
\ee
The second line gives the constraint:
\bea\label{secondsea}
C^\nu_C C^\lambda_Gf^{EFG}{}_D f^{BDC}{}_A=0\;.
\eea
We will see shortly that $f^{ABC}{}_D$ must satisfy a `fundamental identity' for real 3-algebras \cite{Bagger:2006sk,Bagger:2007jr,Bagger:2007vi,Gustavsson:2007vu} and as a result (\ref{secondsea}) will follow from (\ref{Cconstraints}).

We also see that the third line gives the constraint:
\bea
C^\lambda_C D_\lambda X^I_Df^{BDC}{}_A=0\;,
\eea
which implies that the physics is (largely) five-dimensional. In addition,  acting with supersymmetry leads to 
\bea
C^\lambda_C D_\lambda\Psi_Df^{BDC}{}_A=0\;.
\eea

\subsection{Closure on $H_{\mu\nu\lambda \;A}$}

We continue with the closure on the antisymmetric tensor field-strength. In particular we find:
\bea\label{Hxfms}
[\delta_1, \delta_2]H_{\mu\nu\lambda\;A} &=& \;v^\rho D_\rho H_{\mu\nu\lambda\;A}-2i(\bar\epsilon_2\Gamma_\rho\Gamma^I\epsilon_1)C^\rho_CX^I_Dg^{DBC}{}_A H_{\mu\nu\lambda\; B} \cr&&
-6i(\bar\epsilon_2\Gamma_{[\mu}\Gamma^I\epsilon_1)\Big(\tilde F_{\nu\lambda]}{}^C{}_A-C^\rho_BH_{\nu\lambda]\rho\; D}g^{CDB}{}_A\Big)X^I_C\cr
&&-6i(\bar\epsilon_2\Gamma_{\rho[\mu\nu}\Gamma^{IJ}\epsilon_1)C^\rho_BX^I_CD_{\lambda]}X^J_D(f^{CDB}{}_A-g^{CDB}{}_A)\cr
&&-\frac{3i}{8}(\bar \epsilon_2\Gamma^\sigma\Gamma^J\epsilon_1) (\bar\Psi_C\Gamma_{\mu\nu\lambda\rho\sigma}\Gamma^J\Psi_D)C^\rho_B(h^{DBC}{}_A-g^{CDB}{}_A)\cr
&&+2i(\bar\epsilon_2\Gamma^\tau\Gamma^K\epsilon_1) \epsilon_{\mu\nu\lambda\rho\sigma\tau}C^\rho_{B}C^\sigma_E X^I_C X^{I}_F X^{K}_G {g^{D[B|C}}_A{f^{FG|E]}}_D \cr
&&+i(\bar\epsilon_2\Gamma_{\mu\nu\lambda}\Gamma_{LM}\epsilon_1)\epsilon^{IJKLM}C^\kappa_{B}C_{\kappa\;E} X^I_C X^J_F X^K_G {g^{DB[C}}_A{f^{FG]E}}_D \cr
&&+3i(\bar\epsilon_2\Gamma_{\rho[\mu\nu}\Gamma_{LM}\epsilon_1)\epsilon^{IJKLM}C^\rho_{B}C_{\lambda]\;E} X^I_C X^J_F X^K_G {g^{DB[C}}_A{f^{FG]E}}_D \cr
&&
+v^\rho\Big(4D_{[\mu}H_{\nu\lambda\rho]\;A}+\epsilon_{\mu\nu\lambda\rho\sigma\tau}C^\sigma_B X^I_CD^\tau X^I_Dg^{CDB}{}_A \cr
&&\qquad \qquad\qquad+ \frac{i}{2}\epsilon_{\mu\nu\lambda\rho\sigma\tau}C^\sigma_B \bar\Psi_C\Gamma^\tau \Psi_Dg^{CDB}{}_A\Big) \cr
&= & v^\rho D_\rho H_{\mu\nu\lambda\;A}+\tilde \Lambda^B{}_AH_{\mu\nu\lambda\;B}\;,
\eea 
where again we have written the required expression in the final line.

The second term of the first line gives the correct gauge transformation if
\bea
g^{CDB}{}_A = f^{CDB}{}_A\;.
\eea
In addition one sees that the second and third lines now vanish. The fourth line will vanish if
\bea
h^{DBC}{}_A =g^{CDB}{}_A\;.
\eea
Given the previous conditions this implies that $f^{CDB}{}_A=-f^{CBD}{}_A$ and thus $f^{CDB}{}_A$ is totally antisymmetric in $C,D,B$. Just as with multiple M2-branes, consistency of the gauge symmetries $\tilde \Lambda^B{}_A$ implies that the structure constants satisfy the fundamental identity:
\bea
f^{[ABC}{}_Ef^{D]EF}{}_G=0\;.
\eea
Using this, along with the second condition in (\ref{Cconstraints}) one sees that all the terms quadratic in $C^\lambda_A$ vanish.

Demanding that the seventh line vanishes  gives the $H$-equation of motion:
\bea
D_{[\mu}H_{\nu\lambda\rho]\;A}+\frac{1}{4}\epsilon_{\mu\nu\lambda\rho\sigma\tau}C^\sigma_B X^I_CD^\tau X^I_Df^{CDB}{}_A +\frac{i}{8}\epsilon_{\mu\nu\lambda\rho\sigma\tau}C^\sigma_B \bar\Psi_C\Gamma^\tau \Psi_Df^{CDB}{}_A=0\;.
\eea
We  see that the Bianchi identity $D_{[\lambda} \tilde F_{\mu\nu]}{}^A{}_B=0$, along with the $H$-equation of motion, implies that
\bea
C^\rho_C D_\rho H_{\mu\nu\lambda\; D}f^{CDB}{}_A=0\;.
\eea

One could try to introduce a field $B_{\mu\nu\; A}$ such that $H_{\mu\nu\lambda\; A}=3 D_{[\mu}B_{\nu\lambda]\; A}$. This would lead to the algebraic constraint
\bea\label{Beq}
\tilde F_{[\mu\nu}{}^B{}_A B_{\lambda\rho]\; B}+\frac{1}{6}\epsilon_{\mu\nu\lambda\rho\sigma\tau}C^\sigma_B X^I_CD^\tau X^I_Df^{CDB}{}_A +\frac{i}{12}\epsilon_{\mu\nu\lambda\rho\sigma\tau}C^\sigma_B \bar\Psi_C\Gamma^\tau \Psi_Df^{CDB}{}_A=0\; 
\eea
but this over-constrains the fields and hence there cannot exist a suitable $B_{\mu\nu\; A}$. We will verify this in the next section.

\subsection{Closure on $\Psi_A$}

Finally we look at the closure on the fermions. Using the relations that we found above, one gets
\bea\label{closurepsi}
[\delta_1, \delta_2] \Psi_A &=& v^\mu D_\mu\Psi_A +\tilde \Lambda^B{}_A\Psi_B\cr
&&+\frac{3i}{4}(\bar\epsilon_2\Gamma_\sigma\epsilon_1)\Gamma^\sigma(\Gamma^\mu D_\mu\Psi_A+X^I_CC^\nu_B\Gamma_\nu\Gamma^I \Psi_D f^{CDB}{}_A)\cr
&&-\frac{i}{4}(\bar\epsilon_2\Gamma_\sigma\Gamma^K\epsilon_1)\Gamma^\sigma\Gamma^K(\Gamma^\mu D_\mu\Psi_A+X^I_CC^\nu_B\Gamma_\nu\Gamma^I\Psi_D f^{CDB}{}_A)\;.
\eea
Here we achieve closure with the Fermion equation of motion
\bea
\Gamma^\mu D_\mu\Psi_A+X^I_CC^\nu_B\Gamma_\nu\Gamma^I \Psi_D f^{CDB}{}_A=0\; ,
\eea
which agrees with the condition (\ref{fermionanticipated}) that we obtained from demanding that the selfduality of $ H_{\mu\nu\lambda \; A}$ is preserved under supersymmetry.

We can also take a supersymmetry variation of the Fermion equation to obtain the scalar equation of motion:
\bea
D^2 X^I -\frac{i}{2}\bar\Psi_CC^\nu_B\Gamma_\nu\Gamma^I \Psi_D f^{CDB}{}_A - C^\nu_B C_{\nu G} X^J_CX^J_EX^I_F f^{EFG}{}_{D}f^{CDB}{}_A
=0\;.
\eea

\subsection{Summary}
Let us summarise the results of our computation.  The equations
\bea\label{eq1}
0 &=&D^2 X_A^I -\frac{i}{2}\bar\Psi_CC^\nu_B\Gamma_\nu\Gamma^I \Psi_D f^{CDB}{}_A - C^\nu_B C_{\nu G} X^J_CX^J_EX^I_F f^{EFG}{}_{D}f^{CDB}{}_A
 \cr
0 &=& D_{[\mu}H_{\nu\lambda\rho]\;A}+\frac{1}{4}\epsilon_{\mu\nu\lambda\rho\sigma\tau}C^\sigma_B X^I_CD^\tau X^I_Df^{CDB}{}_A + \frac{i}{8}\epsilon_{\mu\nu\lambda\rho\sigma\tau}C^\sigma_B \bar\Psi_C\Gamma^\tau \Psi_Df^{CDB}{}_A \cr
0 &=& \Gamma^\mu D_\mu\Psi_A+X^I_CC^\nu_B\Gamma_\nu\Gamma^I\Psi_D f^{CDB}{}_A \cr
0&=& \tilde F_{\mu\nu}{}^B{}_A - C^\lambda_CH_{\mu\nu\lambda\; D}f^{CDB}{}_A\cr
0 &=& D_\mu C^\nu_A = C^\mu_CC^\nu_Df^{BCD}{}_A \cr
0 &=& C^\rho_CD_\rho X^I_D f^{CDB}{}_A = C^\rho_CD_\rho \Psi_D f^{CDB}{}_A =C^\rho_CD_\rho H_{\mu\nu\lambda\;A} f^{CDB}{}_A \;,
\eea
with $H_{\mu\nu\lambda\;A}$  selfdual, are invariant under the six-dimensional $(2,0)$ supersymmetry transformations
\bea\label{ansatz}
\delta X^I_A &=& i \bar \epsilon \Gamma^I \Psi_A\nn\\
\delta \Psi_A &=& \Gamma^\mu \Gamma^I D_\mu X_A^I \epsilon + \frac{1}{ 3!}\frac{1}{2} \Gamma_{\mu\nu\lambda}H_A^{\mu\nu\lambda}\epsilon- \frac{1}{2}\Gamma_\lambda \Gamma^{IJ} C^\lambda_B X^I_C X^J_D {f^{CDB}}_A\epsilon\nn\\
\delta H_{\mu\nu\lambda\; A} &=& 3 i \bar \epsilon \Gamma_{[\mu\nu}D_{\lambda]} \Psi_A +  i\bar \epsilon \Gamma^I \Gamma_{\mu\nu\lambda\kappa}C^\kappa_{B} X^I_C \Psi_D{f^{CDB}}_A \nn\\
\delta  \tilde A_{\mu\;A}^{\;B} &=& i \bar \epsilon \Gamma_{\mu\lambda} C^\lambda_C \Psi_D {f^{CDB}}_A\\
\delta C^\mu_A  &=& 0\;,
\eea  
provided that $f^{ABC}{}_D= f^{[ABC]}{}_D$ obeys the fundamental identity: $f^{[ABC}{}_Ef^{D]EF}{}_G=0$.

The above are precisely the structure constants for the real 3-algebra of \cite{Bagger:2006sk,Bagger:2007jr,Bagger:2007vi,Gustavsson:2007vu}. Furthermore we need to endow the 3-algebra with an inner product $\Tr (T^A, T^B) = h^{AB}$ with which one can construct gauge-invariant quantities. This in turn implies that $f^{ABCD} = h^{DE}f^{ABC}{}_E$ is antisymmetric in $C,D$ and hence antisymmetric in all of $A,B,C,D$.

\section{Relation to Five-Dimensional SYM}

3-algebras can be classified according to the signature of the metric in group space. In particular there is exactly one \cite{Papadopoulos:2008sk,Gauntlett:2008uf} Euclidean four-dimensional 3-algebra, $\mathcal A_4$, as well as an infinite set of Lorentzian 3-algebras   \cite{Gomis:2008uv,Benvenuti:2008bt,Ho:2008ei}. In this section we move on to investigate the vacuum solutions of our theory for these two possibilities, but one can also consider three-algebras with   more than one timelike directions \cite{deMedeiros:2008bf,Ho:2009nk,deMedeiros:2009hf}.  For a recent review of 3-algebras in physics see \cite{deAzcarraga:2010mr}. 

\subsection{Lorentzian Case }\label{lorentzian}

The Lorentzian 3-algebras can be constructed \eg~as in \cite{Benvenuti:2008bt} by starting with an ordinary Lie algebra $\mathcal G $ and adding two lightlike generators $T^\pm$ such that $A = +,-,a,b,...$, raising the total dimension to $dim(\mathcal G) + 2$. The structure constants are given by
\be\label{lorentzian}
{f^{+ab}}_c = {f^{ab}}_c\;,\qquad {f^{abc}}_- = f^{abc}\;,
\ee
with ${f^{ab}}_c$ the structure constants of the Lie algebra $\mathcal G$ and all remaining components of ${f^{ABC}}_D$ vanishing. The metric is  given by
\be
h_{AB}= \left(\begin{array}{cc|ccc}
0 & -1 & 0 & \dots & 0\\
-1 & 0 & 0 & \dots & 0\\ \hline
0 & 0 &  && \\
\vdots & \vdots & & h_{\mathcal{G}}&\\
0 & 0 &  &&
\end{array}\right)\;.
\ee

We next look for vacua of this theory in the particular case of $\mathcal G = \mathfrak {su}(N)$ by  expanding around a particular point
\be\label{vev}
\langle C^\lambda_A \rangle = g \delta_5^\lambda \delta^+_A \;,
\ee
while all other fields are set to zero. One then has from the fourth line of (\ref{eq1}) that
\be\label{FvH}
\tilde F_{\alpha\beta}{}^b{}_a =  g H_{\alpha\beta 5\; d}f^{db}{}_a\;,
\ee
with  $\mu = \alpha, 5$ and all other components of $\tilde F_{\mu\nu}{}^B{}_A$ zero. As a result the latter correspond to flat connections that can be set to zero up to  gauge transformations and the fifth line of (\ref{eq1})  reduces to $\pd_\mu g = 0$, rendering $g$ constant. 

The rest of (\ref{eq1})  become:
\bea\label{YMeom}
0 &=&\tilde  D^\alpha \tilde  D_\alpha X_a^I - g \frac{i}{2}\bar\Psi_c\Gamma_5\Gamma^I \Psi_d f^{cd}{}_a -  g^2 X^J_cX^J_eX^I_f f^{ef}{}_{d}f^{cd}{}_a \cr
0 &=& \tilde  D_{[\alpha}H_{\beta\gamma] 5\;a}\cr
0 &=& \tilde  D^\alpha H_{\alpha\beta 5\; a}+\frac{1}{2} g f^{cd}{}_a ( X^I_c \tilde  D_\beta X^I_d + \frac{i}{2} \bar\Psi_c\Gamma_\beta \Psi_d ) \cr
0 &=& \Gamma^\mu \tilde  D_\mu\Psi_a+ g X^I_c\Gamma_5 \Gamma^I\Psi_df^{cd}{}_a\cr
0 &=& \pd_5 X^I_d = \pd_5 \Psi_d = \pd_5 H_{\mu\nu\lambda\;d} \;,
\eea
where $\tilde  D_\alpha X^I_a= \pd_\alpha X^I_a - \tilde A_{\alpha}{}^b{}_a X^I_b$, while one also has from (\ref{ansatz}) that
\bea\label{YMsusy}
\delta X^I_a &=& i \bar \epsilon \Gamma^I \Psi_a\cr
\delta \Psi_a &=& \Gamma^\alpha \Gamma^I \tilde  D_\alpha X_a^I \epsilon + \frac{1}{2} \Gamma_{\alpha\beta} \Gamma_5H_a^{\alpha\beta 5}\epsilon- \frac{1}{2}\Gamma_5 \Gamma^{IJ}  X^I_c X^J_d {f^{cd}}_a\epsilon\cr
\delta  \tilde A_{\alpha \;a}^{\;b} &=& i \bar \epsilon \Gamma_{\alpha}\Gamma_5  \Psi_d {f^{db}}_a\;.
\eea  
We immediately see that with the identifications
\be
g= g_{YM}^2\;, \qquad H_{\alpha\beta5}^a = \frac{1}{g_{YM}^2} F_{\alpha\beta}^a\;, \qquad \tilde A_{\alpha\;a}^{\;b} =  A_{\alpha\;c} {f^{cb}}_a\;
\ee
we recover the equations of motion, Bianchi identity and supersymmetry transformations (\ref{SYM}) of five-dimensional $\SU(N)$ super-Yang-Mills  theory. In particular since $g$ has scaling dimension $-1$, we see that $g_{YM}$ also has the correct scaling dimension. Furthermore the fundamental identity reduces to the Jacobi identity for the structure constants of $\mathfrak{su}(N)$. Hence the off-shell $\SO(5,1)$ Lorentz and conformal symmetries are  spontaneously broken to an $\SO(4,1)$ Lorentz invariance.

However we also have the additional equations
\bea\label{20eom}
0 &=&\pd^\mu \pd_\mu X_\pm^I  \cr
0 &=& \pd_{[\mu}H_{\nu\lambda\rho]\;A} \cr
0 &=& \Gamma^\mu \pd_\mu\Psi_A \;,
\eea
with transformations 
\bea\label{20susy}
\delta X^I_\pm &=& i \bar \epsilon \Gamma^I \Psi_\pm\nn\\
\delta \Psi_\pm &=& \Gamma^\mu \Gamma^I \pd_\mu X_\pm^I \epsilon + \frac{1}{ 3!}\frac{1}{2} \Gamma_{\mu\nu\lambda}H_\pm^{\mu\nu\lambda}\epsilon\nn\\
\delta H_{\mu\nu\lambda\; \pm} &=& 3 i \bar \epsilon \Gamma_{[\mu\nu}\pd_{\lambda]} \Psi_\pm \;.
\eea  
These comprise two free, abelian $(2,0)$ multiplets in six dimensions. 

Finally we return to the existence of a 2-form $B_{\mu\nu\;A}$. To start we note that (\ref{Beq}) only acts on the nonabelian fields. For the abelian sector we have $\partial_{[\mu}H_{\nu\lambda\rho]\;\pm}=0$  and hence we can locally write 
$H_{\mu\nu\lambda\;\pm}=3\partial_{[\mu}B_{\nu\lambda]\;\pm}$. 

Next, let us look at the nonabelian fields. From (\ref{FvH}) we have, assuming $D_5B_{\alpha\beta\; a}=0$,
\bea
\tilde F_{\alpha\beta}{}^b{}_a =g(\tilde D_{\alpha} B_{\beta 5\;c}-\tilde D_{\beta} B_{\alpha 5\;c})f^{cb}{}_a\;.
\eea
However we should compare this using $ \tilde F^{\;\;a}_{\alpha\beta\;b} = \pd_\beta \tilde A^{\;a}_{\alpha \;b} - \pd_\alpha \tilde A^{\;a}_{\beta \;b} - \tilde A_{\alpha\;c}^{\;a}\tilde A_{\beta\;b}^{\;c} + \tilde A_{\beta\;c}^{\;a}\tilde A_{\alpha\;b}^{\;c}$ and $\tilde D_\alpha B_{\beta 5\; a}=\partial_\alpha B_{\beta5\; a}-\tilde A_\alpha{}^b{}_aB_{\beta5\; b}$. Examining the derivative terms leads to
\bea
\tilde A_\alpha{}^a{}_b = -gB_{\alpha 5\; c}f^{cb}{}_a\; .
\eea
If we now look at the nonlinear terms we require $f^{fb}{}_cf^{ec}{}_a-f^{eb}{}_cf^{fc}{}_a = 2 f^{ef}{}_df^{db}{}_a$ but using the Jacobi identity one finds instead $f^{fb}{}_cf^{ec}{}_a-f^{eb}{}_cf^{fc}{}_a = f^{ef}{}_df^{db}{}_a$. Thus we conclude that there is no $B_{\mu\nu\; A}$ in general.

To summarise, for the choice of a Lorentzian 3-algebra the vacua of the theory interestingly correspond to the ones for five-dimensional super-Yang-Mills along with two free, abelian $(2,0)$ multiplets which are genuinely six-dimensional.  Presumably one must be gauged away in order to have a well-defined system of equations with positive definite energy.

\subsection{The Euclidean case and `D4 to D4'}

Using the Euclidean 3-algebra is qualitatively rather similar: For the $\mathcal A_4$ 3-algebra  the structure constants coincide with the invariant tensor of $\SO(4)$, $f^{ABCD} = \varepsilon^{ABCD}$. Singling out one of the $\SO(4)$ directions, $A= a,4$ and  expanding the theory around a vev $\langle C^\mu_A\rangle = v \delta^\mu_5 \delta_A^4$ leads to (\ref{YMeom}) and (\ref{YMsusy}), where $f^{abc}$ are now the structure constants of $\SU(2)$ and we can  once again  identify the theory around this vacuum as five-dimensional $\SU(2)$ super-Yang-Mills. In this case one has only a single six-dimensional $(2,0)$ tensor multiplet, obtained by (\ref{20eom}) and (\ref{20susy}) by considering the replacement $(\pm\to 4)$. Thus the $\mathcal A_4$-algebra does not exhibit any qualitative differences compared to the Lorentzian result, in contrast to the case of three-dimensional 3-algebra theories with 16 supercharges.

In fact it would have been possible to arrive at our initial ansatz for the six-dimensional theory by working backwards in the spirit of \cite{Ezhuthachan:2008ch}: Starting with the $\SU(N)$ super-Yang-Mills theory in five dimensions and considering the set of equations of motion and supersymmetry transformations, we rename the YM coupling $g_{YM}^2 \equiv C^5_+$ and the gauge field $ F_{\alpha\beta}^a \equiv \frac{1}{g_{YM}^2} H_{\alpha\beta5}^a$, $ A_{\alpha\;c} {f^{cb}}_a \equiv \tilde A_{\alpha\;a}^{\;b}$. We then promote the coupling into a field, while imposing the external constraint $\pd_\alpha C^5_+ = 0$. This provides the off-shell conformal invariance. Finally we perform a trivial lift to six dimensions (by making the fields six-dimensional but imposing the external constraint that none depend on the new direction), add the free abelian $(2,0)$ tensor multiplets and the flat gauge fields that complete $\tilde F_{\mu\nu}{}^B{}_A$, and use the relations (\ref{lorentzian}) between the Lie and 3-algebra generators. By $\SO(5,1)$-covariantising the resulting equations and writing everything in terms of generic 3-algebra expressions one arrives at (\ref{eq1}) and (\ref{ansatz}). 

Hence, with the use of Lorentzian 3-algebras, it is possible to go  from a conventional description of five-dimensional super-Yang-Mills,  the low-energy theory on the D4-brane worldvolume, to an equivalent 3-algebraic version with off-shell $\SO(5,1)$ and conformal symmetries, as was also the case for D2-branes \cite{Ezhuthachan:2008ch}.

\section{Null Reduction and BPS states}

 It is of interest to investigate whether or not the $(2,0)$ theory derived above  can have any relevance to multiple M5-branes. As we have seen the nonabelian sector of the theory is essentially five-dimensional super-Yang-Mills and therefore more appropriately describes multiple D4-branes. However in this section we will discuss a slightly different choice for $C^\mu_A$. 

In particular let us consider six-dimensional coordinates $x^\mu=(u,v,x^i)$ where $u = \frac{1}{\sqrt{2}}(x^0-x^5)$, $v= \frac{1}{\sqrt{2}}(x^0+x^5)$ and $i=1,2,3,4$. 
 Following the conventions of Section~\ref{lorentzian} we choose any Lorentzian 3-algebra by having that $\langle C^\mu_A \rangle = g\delta^\mu_v\delta^+_A$.  The abelian sector of the theory again consists of free 6-dimensional $(2,0)$ tensor multiplets. However the nonabelian sector is a novel supersymmetric system that effectively lives in $4$ space and $1$ null dimensions with 16 supersymmetries and an $\SO(5)$ R-symmetry. The equations of motion for the nonabelian fields are
\bea
0 &=& D^2 X^I_a-\frac{ig}{2}\bar\Psi_c\Gamma_v\Gamma^I \Psi_d f^{cd}{}_a \cr
0 &=&  \Gamma^\mu D_\mu\Psi_a+gX^I_c \Gamma_v\Gamma^I\Psi_d f^{cd}{}_a   \cr
0 &=& D_{[\mu}H_{\nu\lambda\rho]\;a}-\frac{g}{4}\epsilon_{\mu\nu\lambda \rho \tau v} X^I_cD^\tau X^I_df^{cd}{}_a - \frac{ig}{8}\epsilon_{\mu\nu\lambda\rho\tau v} \bar\Psi_c\Gamma^\tau \Psi_df^{cd}{}_a   \cr
0 &=&  \tilde F_{\mu\nu}{}^b{}_a - gH_{\mu\nu v\; d}f^{db}{}_a 
\eea
with $D_v$ vanishing on all fields. Note that the potential term for the scalar fields vanishes. It would be interesting to try to relate this system to a matrix-model or lightcone description of M5-branes.
 
 These coordinates are well suited for describing the intersection of M2-branes suspended between parallel M5-branes:
\bea
 \begin{array}{cccccccc}
{\rm M5}:  &0 & 1&2&3&4&5  \\
{\rm M2}: &  0&&&&&5&6   \\
 \end{array}
\eea
The resulting solution should appear as a nonabelian version of the selfdual string   \cite{Howe:1997ue}.
The preserved supersymmetries satisfy $\Gamma_{uv}\Gamma_6\epsilon = \epsilon$ in addition to $\Gamma_{uv1234}\epsilon =\epsilon$. In fact, by choosing $C^\mu_A$ to point along the $v$-axis,  we can  use the above equations to describe the right-moving modes of the selfdual string \ie\ modes with $D_v=0$.

The BPS solitons for the abelian fields will comprise of selfdual strings   as well as their `neutral string' generalisations, as studied in \cite{Gauntlett:1998wb}. Thus let us set all the abelian fields to zero here. The constraints imply that $D_v$ vanishes when acting on all the nonabelian fields. 

After setting the fermions to zero the BPS condition is $\delta\Psi_A = 0$, which  becomes:
\bea\label{BPS}
0 &=&D_iX^I_a \Gamma^i\Gamma^I \epsilon- D_u X^I_a \Gamma^I\Gamma^u\epsilon \cr &&+H_{uvi\;a}\Gamma^i\Gamma^{uv}\epsilon+\frac{1}{4}H_{uij\;a}\Gamma^{ij}\Gamma^u\epsilon +\frac{1}{4}H_{vij\;a}\Gamma^{ij}\Gamma^v\epsilon\cr
&&-\frac{g}{2}X^I_cX^J_df^{cd}{}_a\Gamma^{IJ}\Gamma^u\epsilon\ .
\eea
Note that here we have not directly included the contributions from $H_{ijk\;a}$  since it is related by selfduality to $H_{uv i\; a}$. In addition one finds that $H_{vij\;a}$ is antiselfdual and  $H_{uij\;a}$ is selfdual in the transverse space.

The interesting nonabelian solutions should involve a nonvanishing $\tilde F_{ij}{}^b{}_a = H_{vij\;c}f^{cb}{}_a$. For (\ref{BPS}) to be satisfied and the solution to be supersymmetric we need to impose the left-moving projector: $\Gamma^v\epsilon=0$. In fact this projector breaks another half of the remaining supercharges bringing the number of preserved ones to 4. Eq.~(\ref{BPS}) now becomes
\be
D_iX^I_a \Gamma^i\Gamma^I \epsilon -H_{uvi\;a}\Gamma^i\Gamma^{6}\epsilon- D_u X^I_a \Gamma^I\Gamma^u\epsilon +\frac{1}{4}H_{uij\;a}\Gamma^{ij}\Gamma^u\epsilon  = 0\;,
\ee
where we have used the fact that from the second projection, $\Gamma_{uv}\Gamma_6 \epsilon = 0$, one has $\Gamma^{uv} \epsilon= -\Gamma^6 \epsilon$. The first two terms vanish by having that $D_iX^I_a=0$, for $I>6$, and $H_{uv i\;a} = D_i X^6_a$, while the remaining ones after imposing  $D_uX^I_a=H_{uij\;a}=0$.

We summarise the $\frac{1}{4}$-BPS equations for our null-reduced theory:
\bea
H_{uv i\;a} = D_i X^6_a\;,\qquad  H_{vij\;a} = -\frac{1}{2}\epsilon_{ijkl}H_{vkl}{}_{\;a}\; ,
\eea
with 
\be
H_{uvi\;c}f^{ca}{}_b  = - \tilde F_{ui}{}^a{}_b\;,\qquad
H_{vij\; c}f^{ca}{}_b = \tilde F_{ij}{}^a{}_b\;.
\ee
 The solutions to these equations consist of taking a nonabelian four-dimensional instanton $\tilde F_{ij}{}^a{}_b$ along with a solution to 
\bea
D^i D_i X^6_a=0\ ,
\eea
in order to satisfy the $H$-equation of motion.

These are essentially the BPS equations of a ``dyonic-instanton string'' \cite{Lambert:1999ua}, the only difference being that here the dyonic-instanton profile is lightlike. They have smooth finite-energy solutions. Although the M-theory interpretation of our $(2,0)$ tensor multiplet is unclear, it is interesting to see these solutions arise since they have the expected properties of a string-like defect between  parallel M5-branes. Here we see that the right-moving modes of the self-dual string are in one-to-one correspondence with dyonic instantons. We also note that  the possibility of relating dyonic-instantons with `W-Bosons' of  $H_{\mu\nu\lambda\; a}$ was already  mentioned in  \cite{Lambert:1999ua}.

\section{Conclusions}

In this paper we have  constructed a nonabelian on-shell six-dimensional $(2,0)$ tensor multiplet. The result was an interacting system of equations where the gauge structure arises from a 3-algebra. The on-shell conditions are quite restrictive however, and for a spacelike choice of $C^\mu_A$ we essentially obtain a reformulation of the D4-brane theory with conformal and $\SO(5,1)$ Lorentz invariance. We additionally investigated a null choice of $C^\mu_A$ which led to a novel supersymmetric system. These equations are not  apparently obtained by dimensional reduction of ten-dimensional super-Yang-Mills.  It is tempting to speculate that they can be related to a lightcone or matrix-model formulation of the M5-brane. We found that this system had dyonic instanton strings  as the right-moving BPS states of M2-branes suspended between parallel M5-branes.

We could also have considered a timelike choice for $C^\mu_A$. This  leads to a nonabelian supersymmetric system in $5$ spatial directions with $16$ supersymmetries and an $\SO(5)$ R-symmetry. This is the correct symmetry to describe a five-dimensional object in 10 spatial dimensions, {\it e.g.} static 5-branes in 11-dimensions. The equations are essentially just those of a Euclidean D4-brane in 10 spatial dimensions, obtained by dimensionally reducing ten-dimensional Wick-rotated super-Yang-Mills theory to five dimensions. However in our case, since we do not need to Wick rotate, the fermions remain Majorana with the correct number of components. 

Another possible case to study is to set $\langle C^\mu_A\rangle=0$. Here we obtain multiple non-interacting copies of the six-dimensional abelian $(2,0)$ tensor multiplet. Although the gauge field strength $\tilde F_{\mu\nu}^A{}_B$ is now constrained to vanish one could consider compactifications on manifolds which admit non-trivial flat connections.

We note that in our construction the nonabelian two-form $B_{\mu\nu\; A}$ never appears and indeed does not seem to exist. Thus we cannot write down any minimal couplings to $B_{\mu\nu\;A}$. This may be problematic in the quantum theory of M5-branes which is expected to contain states which are minimally coupled, such as the selfdual string. This problem is reminiscent of Ramond-Ramond charges in supergravity which appear as solitonic D-brane states even though the supergravity fields do not couple minimally.  

In terms of applications to M5-branes, our results should be viewed as exploratory. Even if we had achieved complete success in writing down a fully six-dimensional system of equations it would still not be  enough to define the quantum theory without also giving a Lagrangian.  Nevertheless  it is of interest to try and see what structures might be at play. The role of 3-algebras, in particular totally antisymmetric Lie 3-algebras, was not an assumption but rather arose through the demands of supersymmetry. Finally we note Euclidean 3-algebras are often associated with a product gauge group of the form $G\times G$. This suggests a method of realising electric and magnetic states in a local manner by considering a $G\times G$ gauge theory and then identifying the electric states of one copy of $G$ with magnetic states of the other, either as an  explicit projection on the spectrum or through an on-shell relation.

\section*{Acknowledgements}  

We would like to thank Jon Bagger, Dario Martelli, Greg Moore, Sunil Mukhi, Savdeep Sethi and Kostas Skenderis for various discussions and comments.  C.P. would like to thank Rutgers University and CERN for hospitality during the course of this work, while N.L. would like to thank Johns Hopkins University. The authors are partially supported by the STFC rolling grant ST/G000395/1.

\section*{Appendix: Notation, conventions and useful relations}

We  work with 32-component Majorana spinors. The $\Gamma$-matrices acting on the latter are  real and satisfy, $\Gamma_m = -C\Gamma_m C^{-1}$, where $C = \Gamma_0$. The fermions are Goldstinos of the symmetry breaking  $\SO(10,1)\to\SO(5,1)\times\SO(5)$ and, by defining the chirality matrix of $\SO(5,1)$ as  $\Gamma_{012345}$, they and the unbroken supersymmetry parameters satisfy the following chirality conditions 
\bea\label{parity}
\Gamma_{012345} \Psi &=& -\Psi\cr
\Gamma_{012345}\epsilon &=& \epsilon\;.
\eea
Their (anti)commutation relations are
\bea
\{\Gamma_\mu, \Gamma_I\} &=& 0\cr
[\Gamma_{012345}, \Gamma_I]&=& 0\cr
\{\Gamma_{012345},\Gamma_\mu\}&=&0\;,
\eea
where $\mu = 0,...,5$, $I = 6,...,10$. The conjugate spinors are defined with the charge conjugation matrix
\be
\bar \Psi = \Psi^T C
\ee
and for our representation we can choose $C = \Gamma_0$. This makes it antisymmetric $C^T= -C$ and antihermitian $C^\dagger = -C$ with $C^{-1} = - C$. One also has that 
\be\label{charge}
C \Gamma_\mu C^{-1} = - \Gamma^T_\mu\qquad{\rm and}\qquad \{C, \Gamma^I\} =0\;.
\ee

We  make use of the appropriate Fierz identities. These are derived from the 11d Fierz identities by reduction. Starting from the standard expansion
\bea
(\bar \epsilon_2 \chi)\epsilon_1 &=& - 2^{-[\frac{11}{2}]} \Big((\bar \epsilon_2 \epsilon_1) \chi+(\bar \epsilon_2 \Gamma_{m}\epsilon_1)\Gamma^{m} \chi-\frac{1}{2!}(\bar \epsilon_2 \Gamma_{mn}\epsilon_1)\Gamma^{mn} \chi-\frac{1}{3!}(\bar \epsilon_2 \Gamma_{mnp}\epsilon_1)\Gamma^{mnp} \chi\cr
&&+\frac{1}{4!}(\bar \epsilon_2 \Gamma_{mnpq}\epsilon_1)\Gamma^{mnpq} \chi+\frac{1}{5!}(\bar \epsilon_2 \Gamma_{mnpqr}\epsilon_1)\Gamma^{mnpqr} \chi\Big)\;,
\eea
the following combination in eleven-dimensions is
\be
(\bar \epsilon_2 \chi)\epsilon_1 -(\bar \epsilon_1 \chi)\epsilon_2 =-\frac{1}{16}\Big((\bar \epsilon_2 \Gamma_m\epsilon_1)\Gamma^m \chi-\frac{1}{2!}(\bar \epsilon_2 \Gamma_{mn}\epsilon_1)\Gamma^{mn} \chi+ \frac{1}{5!}(\bar \epsilon_2 \Gamma_{mnpqr}\epsilon_1)\Gamma^{mnpqr} \chi \Big)\;,
\ee
where $m=0,...,10$. This is what survives  by only keeping symmetric matrices (including the $C$). By doing the split $\SO(10,1)\to\SO(5,1)\times\SO(5)$ one has that since $\epsilon_1$ and $\epsilon_2$ have the same chirality with respect to $\Gamma_{012345}$ (while $\bar \epsilon_1$ and $\bar \epsilon_2$ the opposite)
the surviving terms must involve odd powers of $\Gamma_\mu$'s. Moreover, the expression is only nonvanishing when $\chi$ has the opposite chirality from $\epsilon_1$ and $\epsilon_2$. One then gets
\bea
(\bar \epsilon_2 \chi)\epsilon_1 -(\bar \epsilon_1 \chi)\epsilon_2 &=&-\frac{1}{16}\Big((\bar \epsilon_2 \Gamma_\mu\epsilon_1)\Gamma^\mu \chi-(\bar \epsilon_2 \Gamma_{\mu}\Gamma^I\epsilon_1)\Gamma^\mu \Gamma^I \chi+ \frac{1}{3!}\frac{1}{2!}(\bar \epsilon_2 \Gamma_{\mu\nu\lambda} \Gamma^{IJ}\epsilon_1)\Gamma^{\mu\nu\lambda}\Gamma^{IJ} \chi \cr
&& + \frac{1}{4!}(\bar \epsilon_2 \Gamma_\mu \Gamma^{IJKL}\epsilon_1)\Gamma^\mu\Gamma^{IJKL} \chi + \frac{1}{5!}(\bar \epsilon_2 \Gamma_{\mu\nu\lambda\rho\sigma} \epsilon_1)\Gamma^{\mu\nu\lambda\rho\sigma} \chi \Big)\;.
\eea
It is possible to translate the last line above in terms of fewer $\Gamma$-matrices with the help of $\epsilon$-tensors. The final answer is 
\be\label{fierz}
(\bar \epsilon_2 \chi)\epsilon_1 -(\bar \epsilon_1 \chi)\epsilon_2 =-\frac{1}{16}\Big(2(\bar \epsilon_2 \Gamma_\mu\epsilon_1)\Gamma^\mu \chi-2(\bar \epsilon_2 \Gamma_{\mu}\Gamma^I\epsilon_1)\Gamma^\mu \Gamma^I \chi+ \frac{1}{3!}\frac{1}{2!}(\bar \epsilon_2 \Gamma_{\mu\nu\lambda} \Gamma^{IJ}\epsilon_1)\Gamma^{\mu\nu\lambda}\Gamma^{IJ} \chi \Big)\;.
\ee

\bibliographystyle{utphys}
\bibliography{lorentzianM5}

\end{document}